\documentstyle[12pt,aps,epsf,epsfig]{revtex}

\begin{document}

\def\half{{1\over 2}}

\title{\small \rm \begin{flushright}
\small{ORNL-CTP-96-08}  \\
\end{flushright}
\vspace{2cm}
\large \bf
A Neutron Scattering Study of Magnetic Excitations \\
in the Spin Ladder (VO)$_2$P$_2$O$_7$ \\
\vspace{0.8cm} }

\author{
A.W. Garrett$^1$, S. E. Nagler$^2$, T. Barnes$^3$
and B.C. Sales$^2$
}

\address{
$^1$Department of Physics,
University of Florida,
Gainesville, FL 32611-0448\\
$^2$
Solid State Division,
Oak Ridge National Laboratory, \\
Oak Ridge, TN 37831-6393  \\
$^3$Theoretical and Computational Physics Section,
Oak Ridge National Laboratory, \\
Oak Ridge, TN 37831-6373,  \\
Department of Physics and Astronomy,
University of Tennessee, \\
Knoxville, TN 37996-1501 \\
}

\date{July 1996}

\maketitle

\begin{center}
{\bf Abstract}
\end{center}

\begin{abstract}
In this letter we report results from inelastic neutron scattering
experiments on powder
samples of vanadyl pyrophosphate, (VO)$_2$P$_2$O$_7$.
We see evidence for three magnetic excitations, at 3.5~meV, 6.0~meV and
14~meV. The intensity of the 3.5~meV mode is strong near $Q=0.8$~\AA$^{-1}$,
consistent
with the one-magnon gap mode reported previously and predicted by the
spin-ladder model. The magnetic scattering at 14~meV may be due
to the top of the one-magnon band or the
two-magnon continuum predicted in the ladder model. The 6.0~meV mode
also peaks in intensity at $Q=0.8$\AA$^{-1}$. This mode has not been
reported previously
and was not anticipated by existing theoretical treatments of the spin ladder.

\end{abstract}
\newpage
\section{Introduction}

The antiferromagnetic insulator (VO)$_2$P$_2$O$_7$ \cite{vopo}
(abbreviated VOPO)
is
widely considered to be the prototypical realization of a Heisenberg
spin ladder.
Heisenberg
spin ladders are interesting theoretically \cite{bdrs}
as intermediaries between 1D Heisenberg
antiferromagnets
and 2D systems.
Theoretical studies of this model have concluded that S=1/2 spin ladders
have a gap if they have an even number of chains, but are gapless  for an odd
number of chains.
This is reminiscent of the Haldane result that
half-integer-spin
chains are gapless but
integer-spin chains have energy gaps.
Studies of hole-doped 2-chain ladders in the t-J model find that strong d-wave
hole pairing takes place
on equal-strength ladders, so analogues of the high-Tc
superconductors might be found
in hole-doped spin ladders.
Additional examples of spin ladders are known, including a
Sr-Cu-O series\cite{SCO}, a
La-Cu-O series\cite{LCO}
and Cu$_2$(C$_5$H$_{12}$N$_2$)$_2$Cl$_4$ \cite{cucl}.
The subject of spin ladders was reviewed recently by Dagotto and
Rice \cite{dr}.

The magnetic properties of VOPO have been interpreted in terms of the 2-chain
Hamiltonian
\begin{equation}
{\rm H} =
{\rm J}_{\parallel} \sum_{<ij>}^{chains} \;
\vec {\rm S}_i \cdot \vec {\rm S}_j
   + {\rm J}_{\perp} \sum_{<ij>}^{rungs} \;
\vec {\rm S}_i \cdot \vec {\rm S}_j \ .
\end{equation}
This model gives an accurate fit to the VOPO magnetic
susceptibility \cite{br} with the parameters
J$_{\parallel}$ = J$_{\perp}$ = 7.8  meV.
Theoretical studies of the ladder model predict an energy gap of
E$_{gap}=0.5037$J \cite{wns}, hence
the susceptibility results imply E$_{gap}=3.9$ meV.
A pulsed neutron scattering experiment on VOPO powder \cite{Roger}
observed a gap of 3.7(2) meV,
consistent with this expectation.

Although these aspects of VOPO are consistent with the Hamiltonian (1),
many unanswered
questions remain concerning the dispersion relations,
spectral weight, and detailed band structure of the
magnetic excitations.
We have carried out triple-axis inelastic neutron scattering studies of VOPO
to investigate these issues in more detail.
In addition to confirming the existence of the gapped spin wave mode observed
previously, our results reveal a new and unexpected
magnetic excitation.

The spectrum of
low-lying spin waves in the 2-chain ladder model
has been discussed at length in the literature
\cite{bdrs,br,scdisp}.
The one-magnon band can be approximately described by the function \cite{br}
\begin{equation}
\omega(k) =
\Big[
\omega(0)^2 \cos^{2}(k /2)
+
\omega(\pi)^2 \sin^{2}(k /2)
+c_0^2 \sin^{2}(k ) \Big]^{1/2}
\end{equation}
where $k$ is the one-dimensional wavevector in units in which the
intrachain ion separation is set to unity.
This dispersion relation for
the one-magnon band $\omega(k)$ is shown in Fig.1 together with
the lower edges of the
two-magnon and three-magnon continua. (These are the lowest-energy sums
$\omega(k_1) + \omega(k_2)$ and $\omega(k_1) + \omega(k_2)+ \omega(k_3)$
with the constraint $\sum_i k_i = k$.)
The value J=7.0 meV was
chosen to match the energy gap observed in the present work,
as discussed below, and our dispersion relation
parameters for ${\rm J}_\perp = {\rm J}_\parallel = {\rm J}$
are $\omega(0)=1.72$J,
$\omega(\pi) = 0.5037$J and
$c_0=1.38$J; these values were motivated by finite lattice results.

The lowest-lying one-magnon excitation occurs at the one-dimensional
antiferromagnetic point $k=\pi$, with an energy of
$\omega(\pi)=$ E$_{gap}$. Exact calculations of the
dynamical structure factor $S(k,\omega)$ on a $2\times 8$ lattice
\cite{br,jose}
indicate that the spectral weight of
the one-magnon band peaks strongly near $k=\pi$, so this should be the
most prominent feature seen in neutron scattering.

The two-magnon states span a continuum of levels, beginning at
2E$_{gap}$ at $k=0$. As seen in Fig.1, the lower edge of the two-magnon
continuum
crosses the one-magnon band at an intermediate $k\approx 0.3 \pi$, and for
smaller
$k$ these two-magnon states are the lowest-lying excitations.
Similarly there is a three-magnon continuum, also shown in Fig.1, which
contains
the
first
$k=\pi$ excitation above E$_{gap}$
in the ladder model; this three-magnon continuum begins at 3E$_{gap}$.

Numerical calculations of the structure
factor $S(k,\omega)$ \cite{jose} indicate
that the two-magnon states have moderate spectral weight
near $k=\pi$, at E $\approx 2.2 $J ($\approx 15$ meV in VOPO), and the
one-magnon band may also be observable near $k=0$,
at E $\approx 1.7 $J ($\approx 12 $~meV~in~VOPO). Both excitations have
$S(k,\omega)\approx 0.1\, S(k=\pi,\omega={\rm E}_{gap})$ of the
gap mode.
The $S(k,\omega)$ of the three-magnon continuum at $k=\pi$,
E $\approx 1.5$J ($\approx 11
$ meV) is about a factor of five weaker than these higher excitations.

\section{Experimental Details}

To prepare (VO)$_2$P$_2$O$_7$, stochiometric amounts of V$_2$O$_5$
(44.1 grams, 99.99\%) and NH$_4$H$_2$PO$_4$ (55.9 grams, 99.9\%)
powder were mixed and loaded into a  platinum
crucible. The mixed powders were heated to 500$^{\circ }$C at 1$^{\circ
}$C/minute
and held at 500$^{\circ }$C for 6 hours.
This initial heating allowed the ammonia and water to escape from the
crucible during the decomposition of NH$_4$H$_2$PO$_4$ to P$_2$O$_5$.  The
prereacted
mixture was heated further to a temperature of 1100$^{\circ }$C
and the resulting
liquid was allowed to homogenize for several hours. The platinum crucible
was removed from the furnace at 1100$^{\circ }$C and the liquid was ``cast''
into a
cold, shallow platinum crucible measuring 10 x 4 x 2 cm$^3$. The rapid cooling
of the V$_2$O$_5\cdot$P$_2$O$_5$ liquid resulted in the formation of a
homogenous glass
with a dark green to black color. The above steps were all performed
in air.

The shallow platinum crucible containing the V$_2$O$_5\cdot$P$_2$O$_5$  glass
was then
placed
in a tube furnace in which the oxygen partial pressure could be
controlled. A mixture of 0.1\% oxygen and 99.9\% argon was passed
through the furnace at 40
cc/minute, and this rate and oxygen partial pressure
were maintained during the entire crystal growth procedure. An independent
oxygen sensor was used to monitor the oxygen partial pressure in
the furnace. The V$_2$O$_5\cdot$P$_2$O$_5$ glass was heated to 1050$^{\circ
}$C,
maintained at that temperature for 6 h,
and then cooled at
1$^{\circ }$C/hr to 480$^{\circ }$C, after which the furnace was turned off.
The resulting material consisted of several hundred mm sized single
crystals of (VO)$_2$P$_2$O$_7$ embedded in a polycrystalline matrix of
(VO)$_2$P$_2$O$_7$  and a small amount ($\approx 5$-$15\% $) of an additional
phase, tentatively identified as  V(PO$_3$)$_3$ by neutron
powder diffraction \cite{phase2}.

     Inelastic neutron scattering measurements were carried out using the
HB1A and HB3 triple-axis spectrometers at the
High Flux Isotope Reactor at Oak Ridge National Laboratory \cite{YFB}.
The sample consisted of 18 grams of finely ground VOPO powder sealed
in a 0.5" diameter thin-walled Al cylinder
under a helium
atmosphere.
For the initial
experiment on the HB1A spectrometer the sample was mounted in a standard
He closed-cycle refrigerator with a temperature range of 10K to 300K.
A double-crystal pyrolytic graphite
(PG) (002) monochromator (M) was used to provide an incident
neutron beam with a fixed energy of 14.7 meV. The incident beam intensity was
monitored using a fission counter, and contamination of the beam
by higher-order Bragg reflections was removed using the standard PG filter
method.  Neutrons scattered by the sample (S) were reflected by a PG(002)
analyzer (A) into a $^3$He detector (D). Effective beam collimations
pre-M, M-S, S-A and A-D
of 40$'$-40$'$-40$'$-70$'$ or 40$'$-20$'$-20$'$-70$'$ were used.

A second experiment utilizing the HB3 spectrometer employed PG(002)
monochromator and analyzer crystals with a fixed scattered
neutron energy of 14.7 meV. The collimation was 40$'$-40$'$-40$'$-120$'$,
and the scattered beam was passed through a PG filter. For this experiment
the sample was mounted in a pumped He cryostat capable of reaching~1.5K.

VOPO has a slightly monoclinic unit cell which has lattice parameters
$a~=~7.728~{\rm \AA}$, $b~=~16.589~{\rm \AA}$, $c~=~9.580~{\rm \AA}$,
and $\beta~=~89.98^{\circ }$
at T=296K \cite{latt}.
The ladder is composed of S$=1/2$ V$^{4+}$ ions in -(VO)-(VO)-
chains along the $a$ axis and
V ${ {\rm -O-}\atop {\rm -O-}}$ V rungs oriented along the $b$ axis. The
average
V-V distance along the chain is approximately equal to $a/2$.
Therefore in VOPO the one-dimensional antiferromagnetic points
$k=n\pi$ (with $n$ odd) occur in sheets with $Q_{a}=nQ_{\pi}$,
where $Q_{\pi}=0.8134$ \AA$^{-1}$.

In scattering from {\em powder}
the experimental momentum transfer to the sample $Q$~$=$~$|\vec Q \, |$
does not correspond to the one-dimensional momentum $k$ of the excitations
observed. Instead $k$ is the projection of $\vec Q$ on the ladder axis,
and in the powder one averages over all orientations. Therefore
for a given $Q$ one can excite all ladder modes with the observed
energy transfer and $k \leq Q$.
In particular, one expects the gap mode, which has $k=\pi$,
to appear first at $Q=Q_\pi$
as $Q$ is increased,
and to persist to higher $Q$ due to the powder average.
The scattering intensity observed at $(Q,\omega)$ involves the density
of states as well as $S(k,\omega)$.

\section{Results}

Figure 2 shows the results of constant-$Q$ scans at
$Q=Q_{\pi}$, taken at HB1A.  The data have been scaled to a
fixed monitor value and have been corrected for the
constant-E$_{i}$ kinematical factor $k_{f}\cot\theta_{A}$.
The data at 10.8K clearly show two peaks in the scattering.
The peak intensities are drastically reduced at higher temperatures,
as illustrated
by the data at 41K, which provides
strong evidence that they are both magnetic in origin.
The lower peak at 3.48(3) meV is interpreted as
the one-magnon gap mode previously reported at 3.7(2) meV by
Eccleston {\it et al.}  \cite{Roger}. Their slightly higher energy
may arise from their broader $Q$ acceptance.
The 3.5 meV peak
observed in the present work  has a FWHM significantly greater than the
instrumental resolution limit of 0.7 meV.

The mode at E $=6.0$ meV has not been reported previously.
It appears narrower in energy than the 3.5 meV mode, with
a FWHM only slightly larger than the instrumental
resolution; this suggests that the band is rather flat,
possibly corresponding to a
relatively localized mode.
Comparison with the theoretical magnon bands in Fig.1
shows that this excitation is not
predicted by the simple Heisenberg spin-ladder model (1).

Figure 3 shows the $Q$ dependence of the intensities of the 3.5 meV and 6.0 meV
modes.
Both modes show very similar behavior, with an onset and peak at $Q=Q_{\pi}$.
This is
precisely what one would expect for the lowest-energy one-magnon gap mode.
The fact that the 6.0 meV mode peaks at a value of $Q$ characteristic of VOPO
provides
strong evidence that it actually is a VOPO excitation, and does not arise from
a contaminant phase in the sample.  Both modes also show a slight increase
in intensity in the vicinity of $Q=2Q_{\pi}$, and a gradual falloff in
intensity
at higher $Q$.
The scattering intensity from
a powder sample is not easily interpreted beyond the first Brillouin zone,
but the diminishing intensity with increasing momentum transfer is expected
because of the reduction of the magnetic form factor at large $Q$.
For the V$^{4+}$ ion in VOPO, $|f^{mag}(3Q_{\pi})|^{2} \approx
0.5|f^{mag}(Q_{\pi})|^{2}$.

Constant $Q$ scans at $Q=3Q_{\pi}$ and $Q=5Q_{\pi}$ at a temperature of 1.6K
are presented in the lower panel of Fig.4.
The scans were carried out at HB3 with fixed scattered neutron energy,
which allowed measurements
over a wide range of energy transfers.  The scattering at both wavevectors
exhibits
a broad maximum near 13 meV.
The origin of this broad scattering is unknown, but
since magnetic scattering at $Q=5Q_{\pi}$ is expected to be weak,
it is probably
not magnetic in origin.  Unfortunately, kinematic restrictions did not allow
the same energy scan at lower $Q$, particularly at $Q=Q_{\pi}$ where the
magnetic features should be most prominent.  Inspection of the lower panel of
Fig.4 shows that the $3Q_{\pi}$ scan has
features which were not observed in the $5Q_{\pi}$ scan.
In repetitions (not shown) of the $3Q_{\pi}$ scan at T=45K and T=100K the extra
features are not clearly visible,
and only the broad maximum remains.
The $Q$ and T dependence of the scattering
strongly suggests that the additional
features seen at $3Q_{\pi}$
arise from magnetic excitations.
To investigate this possibility further
the $Q=5Q_{\pi}$ intensity was subtracted from the $Q=3Q_{\pi}$ intensity on
a point-by-point basis;
the result is shown in the upper panel of Fig.4.  The solid line is a fit
of this difference spectrum to a
sum of three
Gaussians and a constant background, revealing features near 3.6~meV and
6.5~meV,
approximately consistent with
the magnetic excitations observed at $Q_{\pi}$. The fit also shows
a higher-energy peak at 14.3 meV.
Eccleston {\it et al.} \cite{Roger}
also saw indications of enhanced scattering near 15 meV at $Q=0.8$ \AA$^{-1}$,
although the details were unclear.

Since 14 meV is close to the predicted onset of the
$k=\pi$ two-magnon continuum as well as the maximum energy of the one-magnon
band (see Fig.1) in the ladder model, it is
tempting to associate the magnetic scattering with
either or both of these features.
However, a detailed comparison of experiment with
the excitation spectrum predicted by (1) is problematical,
since it did not anticipate the 6 meV mode.

\section{Discussion}

In this experiment we have found clear evidence for two low-lying magnetic
excitations in VOPO, at energies of approximately 3.5 meV and 6.0 meV, and
some additional magnetic scattering near 14 meV. The low-lying excitations
peak strongly at $Q=0.8$~\AA$^{-1}$, near the $k=\pi$ point in VOPO, which
suggests that these are $k=\pi$ modes of the VOPO
spin ladder. Although the 3.5 meV
mode and the 14 meV scattering are similar to expectations in the spin-ladder
model, the presence of two low-lying modes at $k=\pi$ is unexpected, and
appears to argue against the validity of the
simple spin-ladder model (1) for VOPO.

Of course one possibility is that the theoretical predictions we associate with
the
spin ladder, such as the spectrum and continua shown in Fig.1, are inaccurate.
Although it appears highly unlikely that {\it two}
low-lying excitations arise at $k=\pi$ in
the model, these
conclusions are based on Lanczos studies of relatively
small clusters (up to $2\times 12$), and careful theoretical investigations
of larger spin ladders would be useful.

Assuming that the theoretical interpretation of the model is accurate, we
conclude that there are magnetic
excitations in VOPO which are not consistent with the Heisenberg
spin-ladder Hamiltonian (1).
In view of this apparent disagreement we should recall the assumptions
which led to
(1) and determine whether they are justified.
The Heisenberg model assumes that each magnetic ion
can be treated as a localized
spin with S$=1/2$, interacting with neighboring ions through an
isotropic interaction.
Whether or not this is true in practice
depends on the ground state and low-lying
excited states of the magnetic ion.

An isolated V$^{4+}$ ion in an octahedral field,
as is approximately realized by the O octahedra in VOPO,
has a $t_{2g}$ triplet ground state \cite{AB};
this triplet is split by the spin-orbit interaction and
by departures from octahedral symmetry, including the formation of (VO)
units along the chains \cite{latt}. For isolated V$^{4+}$ ions in Al$_2$O$_3$
the single-ion excited $t_{2g}$ levels are known to lie just 3.5 meV and
6.6 meV above the ground state \cite{JR},
comparable to the energy scales of the magnetic
modes observed in VOPO.

This raises the possibility that one or more of the magnetic modes observed
in VOPO could be crystal field levels,
which might explain the peak
at 6 meV.
There is experimental evidence against this possibility, however;
a single-ion excitation should not display a strong dependence of the
scattering intensity on $Q$ (other than the magnetic form factor).
Instead we have observed a strong
peak at $Q_\pi $
in the constant-E scan (Fig.3), which
suggests that the 6 meV excitation is
part of a band of delocalized ladder excitations.
Of course this band could involve
excited $t_{2g}$ orbital levels,
in which case VOPO
should be described by a multiband Hubbard model with
$t_{2g}$ V$^{4+}$ orbitals interacting
through intermediate O levels.

Although low-lying $t_{2g}$ V$^{4+}$ orbitals may indeed be present in VOPO,
magnetic susceptibility measurements argue against their presence at
these low
energy scales.
The observed high temperature limit of the susceptibility (known to about 350K
\cite{vopo}) is consistent with
that expected for paramagnetic spins with S=1/2 and g=2. The presence
of additional $t_{2g}$
degrees of freedom would give a much larger susceptibility
as the temperature approached the orbital gap. Thus there is no indication
of excited $t_{2g}$ levels in VOPO to $\approx 30$~meV.
Further experimentation, including infrared spectroscopy and EPR measurements,
is desirable to locate the excited $t_{2g}$ orbital levels of V$^{4+}$ in VOPO.

A more likely possibility is that VOPO actually can be modelled accurately
as an S$=1/2$ spin ladder at low energy scales
(hence the susceptibility is approximately
correct), but there are important additional spin interactions that split the
one-magnon band into two bands, which we observe at 3.5 meV and 6.0 meV at
$k=\pi$.
There are various possibilities for such terms, including ladder-ladder
interactions,
chain dimerization, departures from isotropy,
next-nearest-neighbor interactions and so forth. Two examples of such
interactions
are well known.
Planar anisotropy in the spin Hamiltonian could lead to different gaps for
``in-plane'' and ``out-of-plane'' spin-wave modes;
this has been observed
in S$=1$ antiferromagnetic chain materials \cite{Golinelli}.
A large next-nearest-neighbor interaction
is believed to be present in the CaV$_4$O$_9$
spin lattice \cite{cvo}; similar interactions may be present in VOPO as well.

In conclusion, it appears that a generalization of the simple
Heisenberg spin-ladder Hamiltonian (1)
will be required
to describe the magnetic properties of VOPO.
The correct interpretation of the magnetic excitations would be
facilitated by
more detailed experimental studies of the low-lying bands,
in particular by inelastic
neutron scattering studies of VOPO single crystals. The technical difficulties
of
growing adequate single crystals of
VOPO have precluded such studies to date.
We hope to carry out such an experiment in the near future.

\newpage
\acknowledgements

We would like to acknowledge useful communications with
E.Dagotto, R.S.Eccleston, D.C.Johnston, A.Moreo, J.Riera and C.Torardi.
We thank J.Zarestky and J.Fernandez-Baca for valuable assistance with HB1 and
HB3. Expert technical assistance was provided by S.Moore and G.B.Taylor.

This work was supported in part by the United States Department
of Energy under contract DE-FG05-96ER45280 at the University of Florida, and by
Oak Ridge National Laboratory, managed for the U.S. D.O.E.
by Lockheed Martin Energy Research Corporation under contract
DE-AC05-96OR22464.

\newpage

\begin{center}
{\Large Figure Captions}
\end{center}

\begin{figure}
{Figure~1.
Predicted spin waves in (VO)$_2$P$_2$O$_7$
with J$_\perp = $ J$_\parallel = 7.0$ meV. The one-magnon band
(solid) and the onset of the
two-magnon and three-magnon continua are shown.
}
\end{figure}

\begin{figure}
{Figure~2.
Inelastic scattering from (VO)$_2$P$_2$O$_7$ at $Q=0.8134$ \AA$^{-1}$,
at temperatures of T=10.8K and 40.8K, measured at HB1A with
E$_{i}=14.728$ meV and collimation 40$'$-20$'$-20$'$-70$'$.
The gap mode at 3.5 meV and the 6.0 meV mode are clearly visible in the
10.8K data.  The corrected
neutron counts (see text) are normalized to counts observed in 1250
monitor units; each monitor unit corresponds to about one second of
counting time.
}
\end{figure}

\begin{figure}
{Figure~3.
Constant energy scans of scattering intensity versus momentum transfer $Q$
at $\omega=3.5$ meV (upper panel) and $\omega=6.0$ meV
(lower panel),
normalized to 500 monitor units.
The dashed vertical lines show the values of $Q=nQ_{\pi}$
corresponding to one-dimensional nuclear
($n$ even) and magnetic ($n$ odd) zone centers.
Note the strong peaking at $Q=Q_{\pi}$ in both cases.
Measurements were done at HB1A with fixed incident energy  E$_{i}=14.728$
meV and
collimation 40$'$-40$'$-40$'$-70$'$.
}
\end{figure}

\begin{figure}
{Figure~4.
Constant $Q$ scans at $Q=3Q_{\pi}$ and $Q=5Q_{\pi}$ at
T=1.6K(lower panel) and the subtracted difference scattering (upper panel).
The scattering intensity is normalized to 100 monitor units.
The line in the lower panel is a guide to the eye through the $5Q_{\pi}$ data.
The solid line in the upper panel is a least-squares fit of the difference to a
sum of three Gaussians and a constant background,
with fitted centers equal to 3.62(13)~meV,
6.46(16)~meV and 14.32(22)~meV.  The measurements were carried out at HB3 using
a fixed scattered neutron energy E$_{f}=14.7$ meV and collimation
40$'$-40$'$-40$'$-120$'$.
}
\end{figure}

\newpage

\begin{figure}
$$\epsfxsize=6truein\epsffile{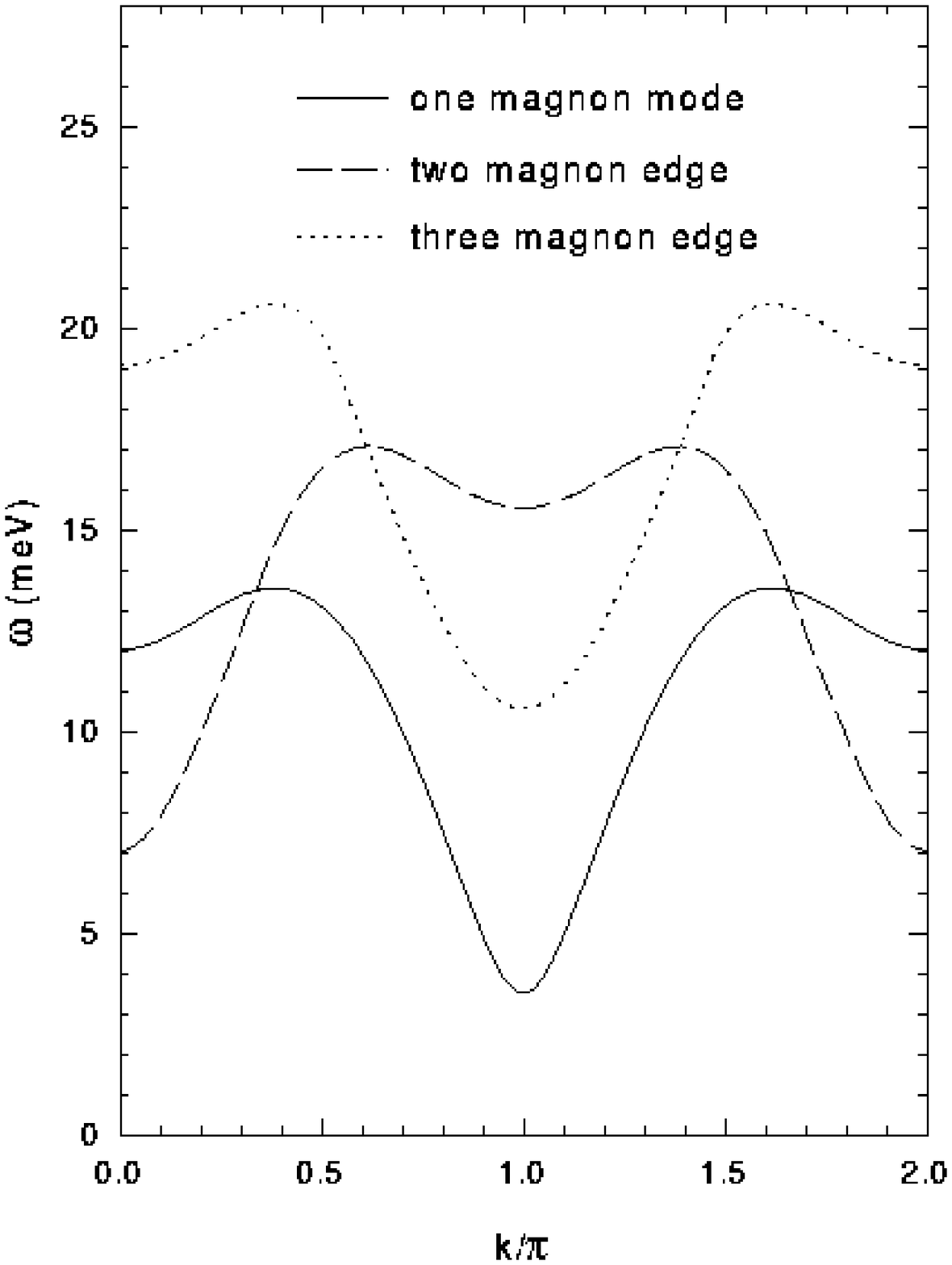}$$
\end{figure}
\center{\Large {Figure 1.}}

\newpage

\begin{figure}
$$\epsfxsize=6truein\epsffile{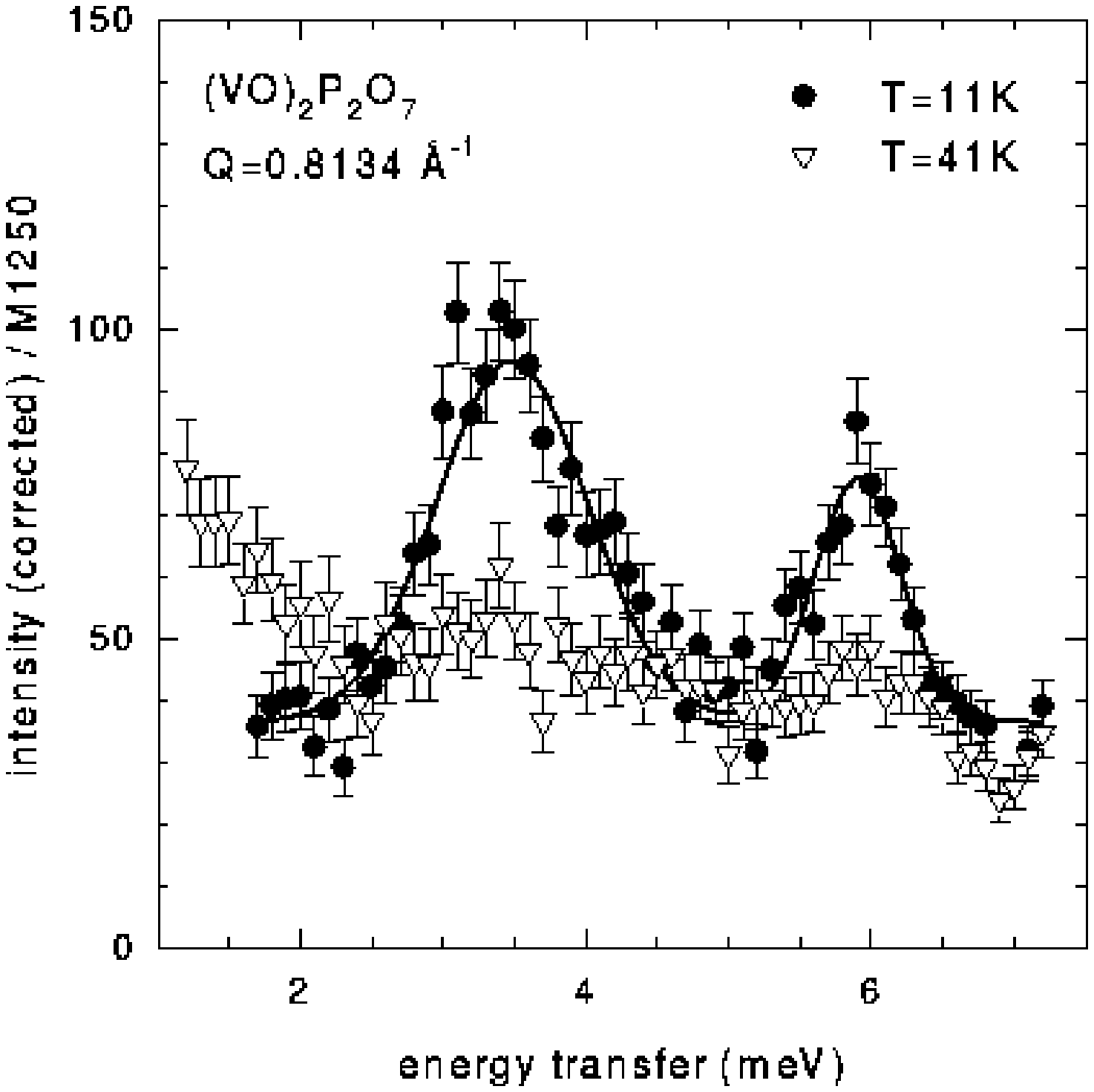}$$
\end{figure}
\center{\Large {Figure 2.}}

\newpage

\begin{figure}
$$\epsfxsize=6truein\epsffile{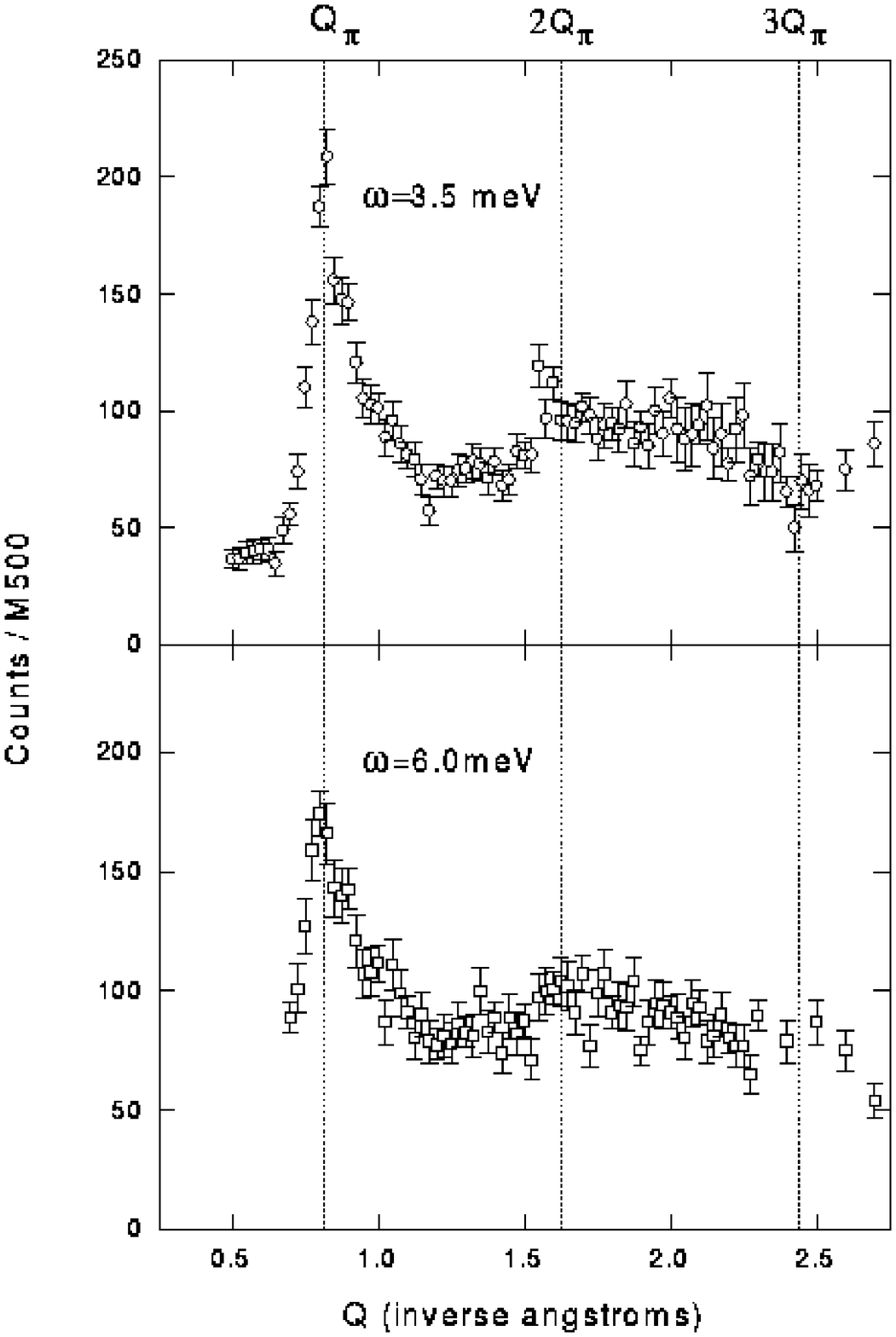}$$
\end{figure}
\center{\Large {Figure 3.}}

\newpage

\begin{figure}
$$\epsfxsize=6truein\epsffile{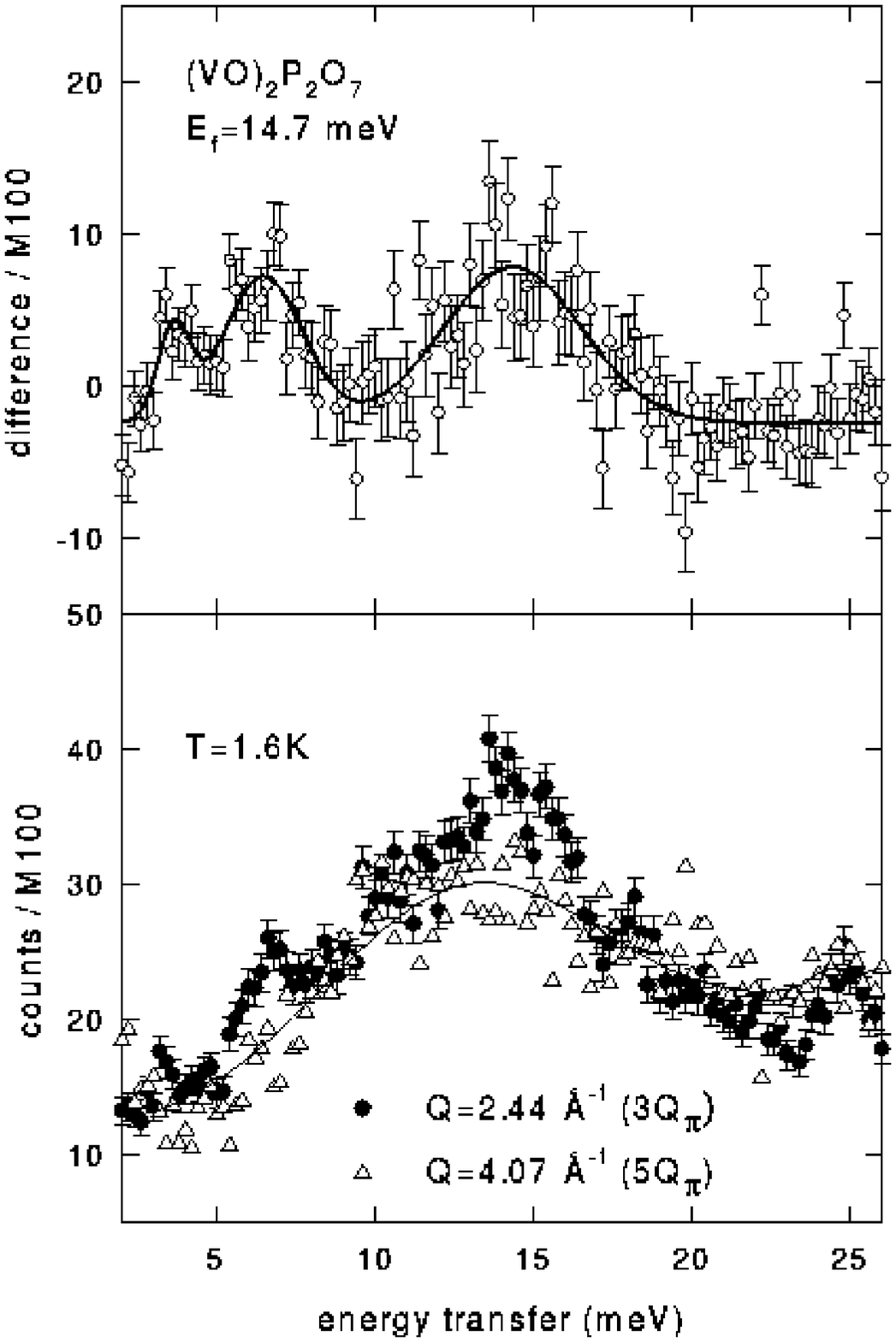}$$
\end{figure}
\center{\Large {Figure 4.}}

\end{document}